\documentclass[a4paper,10pt]{article}

\usepackage{subfig}
\usepackage{graphicx}
\usepackage{cite}
\usepackage{epsfig,amsmath,amssymb,verbatim,mathrsfs,array,layout,textcomp,amssymb,latexsym}

\newcommand{\beq}{\begin{eqnarray}}
\newcommand{\eeq}{\end{eqnarray}}
\newcommand{\be}{\begin{eqnarray}}
\newcommand{\ee}{\end{eqnarray}}

\def\beqa{\begin{eqnarray}}
\def\eeqa{\end{eqnarray}}
\def\bea{\begin{eqnarray}}
\def\eea{\end{eqnarray}}

\newcommand{\bv}{\left(\begin{array}{c}}
\newcommand{\ev}{\end{array}\right)}
\newcommand{\bmtwo}{\left(\begin{array}{cc}}
\newcommand{\bmthree}{\left(\begin{array}{ccc}}
\newcommand{\emn}{\end{array}\right)}
\newcommand{\bmtwoc}{\left\{\begin{array}{cc}}
\newcommand{\bmthreec}{\left\{\begin{array}{ccc}}
\newcommand{\emnc}{\end{array}\right\}}
\newcommand{\ba}{\begin{array}}
\newcommand{\ea}{\end{array}}

\def\lsim{\mathrel{\rlap{\lower4pt\hbox{\hskip1pt$\sim$}}
     \raise1pt\hbox{$<$}}}         
\def\gsim{\mathrel{\rlap{\lower4pt\hbox{\hskip1pt$\sim$}}
     \raise1pt\hbox{$>$}}}         

\begin{document}

\begin{titlepage}

\vskip1.5cm
\begin{center}
  {\Large \bf The three jewels in the crown of the LHC}
\end{center}
\vskip0.2cm

\begin{center}
Yosef Nir\\
\end{center}
\vskip 8pt

\begin{center}
{ \it Department of Particle Physics and Astrophysics,\\
Weizmann Institute of Science, Rehovot 7610001, Israel} \vspace*{0.3cm}

{\tt   yosef.nir@weizmann.ac.il}
\end{center}

\vglue 0.3truecm

\begin{abstract}
  \vskip 3pt \noindent
The ATLAS and CMS experiments have made three major discoveries: The discovery of an elementary spin-zero particle, the discovery of the mechanism that makes the weak interactions short-range, and the discovery of the mechanism that gives the third generation fermions their masses. I explain how this progress in our understanding of the basic laws of Nature was achieved.
\end{abstract}

\end{titlepage}
It is often stated that the Higgs discovery is ``the jewel in the crown" of the ATLAS/CMS research. We would like to argue that ATLAS/CMS made (at least) three major discoveries, each of deep significance to our understanding of the basic laws of Nature:
\begin{enumerate}
\item {\bf The discovery of an elementary spin-0 particle}, the first and only particle of this type to have been discovered.
\item {\bf The discovery of the mechanism that makes the weak interactions short-ranged}, in contrast to the other (electromagnetic and strong) interactions mediated by spin-1 particles.
\item {\bf The discovery of the mechanism that gives masses to the three heaviest matter (spin-1/2) particles}, through a unique type of interactions.
\end{enumerate}
These three breakthrough discoveries can be related, in one-to-one correspondence, to three distinct classes of measurements:
\begin{enumerate}
\item The Higgs boson decay into two photons.
\item The Higgs boson decay into a $W$- or a $Z$-boson and a fermion pair, and the Higgs production via vector boson ($WW$ or $ZZ$) fusion.
\item The Higgs boson decay into fermion pairs, and the Higgs boson production in association with top-antitop pair.
\end{enumerate}

\section{The first jewel: An elementary spin-0 particle}
An elementary particle is a point-like particle, with no inner structure. Until 2012, the list of elementary particles consisted of two broad classes: Matter particles and force carriers. The matter particles are all spin-1/2 (fermions), and there are twelve of them known to us. All structures in the Universe are made of three of these twelve: the electron, the up quark and the down quark. (In a somewhat simplifying language, two up quarks and a single down quark make the proton, and two down quarks and a single up quark make the neutron. Protons, neutrons and electrons make all atoms.) All force carriers have integer spin  (bosons). Four spin-1 force carriers are known to us: the photon (carrier of the electromagnetic force), the gluon (carrier of the strong force), and the $W$- and $Z$-bosons (carriers of the weak force). The gravitational force is presumably mediated by a spin-2 force carrier, the graviton.

In 2012, the ATLAS and CMS experiments announced the discovery of a new particle \cite{Aad:2012tfa,Chatrchyan:2012xdj}. While the discovery was achieved by combining several decay modes, the dominant one was the decay into two photons. Such a decay is possible only for a spin-0 or spin-2 particle. There are two more steps needed to be taken by experiments: First, decide between the spin-0 and spin-2 options and, second, to test the possibility that this new particle is a composite, rather than an elementary, particle.

To decide between the two spin options, a more complex study than just measuring decay rates was needed. The spin of the parent particle affects the angular distributions of the daughter particles coming from its decay. Studies of this question started immediately after the discovery. (For early studies, see for example \cite{Aad:2013xqa,Khachatryan:2014ira}.) These studies showed unambiguously that the newly discovered particle is spin-0. The first -- and the only one to date -- of this kind.

The ways in which this particle is produced and the ways in which it decays call for its identification with the only particle that was predicted by the Standard Model of particle physics and had not been observed until the 2012 discovery -- the Higgs boson \cite{Englert:1964et,Higgs:1964pj}. From here on, we will call the field related to this particle the BEH field.

If the Higgs boson is a composite spin-0 particle, then there should be a whole series of new composite particles, in particular spin-1 particles whose mass scale is (roughly) inversely proportional to the distance scale which characterizes its internal structure. One can test the question of whether the Higgs boson is elementary or composite in three ways. First, an indirect one: Virtual effects of these heavy spin-1 particles would modify various properties of the $W$- and $Z$-bosons. Part of the legacy of the LEP experiments, which operated in CERN between 1989 and 2000, is a large class of precision measurements of these properties. There is no sign of compositeness, and consequently a lower bound on the mass scale of new particles and, equivalently, an upper bound on the distance scale of compositeness, can be extracted. The other two ways are pursued by the LHC experiments: First, the direct search for the new spin-1 particles. No such particles have been discovered to date. Second, precision measurements of various properties of the Higgs boson itself, which would differ if it were elementary or composite. No deviations from the properties of an elementary Higgs boson have been observed. The three ways are comparable in their present reach of testing compositeness. The Higgs boson shows no signs of internal structure down to a scale of $10^{-19}$ meters, some four orders of magnitude below the size of the proton.

\section{A second jewel: Why the weak interaction is short-range}
The gravitational, electromagnetic and strong interactions are mediated by massless mediators -- the graviton, the photon and the gluon. Consequently, they  are long-range. (Color confinement, the phenomenon that quarks and gluons cannot be isolated, renders the long range effects of the strong interaction unobservable.) In contrast, the weak interactions are mediated by massive mediators -- the $W^\pm$ and $Z^0$ bosons -- with masses of order a hundred times that of the proton. Consequently, at distances larger than $10^{-18}$ meters, the weak force is exponentially suppressed.

A common feature of the electromagnetic, strong and weak force is that their mediators are all spin-1. This type of interactions is very special. Our current quantum field theories can predict the existence of this type of interactions, and many of their features, by assuming that Nature has certain symmetries. There are numerous predictions stemming from these symmetries that have been successfully tested by experiments. Yet, these symmetries predict, at least naively, that the spin-1 force carriers should be massless. So, while the symmetry that predicts the electromagnetic interaction also explains why its force carrier, the photon, is massless, and the symmetry that predicts the strong interaction also explains why its force carrier, the gluon, is massless, the symmetry principle that predicts the weak interaction is challenged by the experimental fact that its force carriers, the $W$ and $Z$ bosons, are massive.

This conundrum has a possible solution if a symmetry is respected by the quantum field theory but not by the ground state of the Universe. The theory loses nothing of its predictive power, but the predictions are different from those that would follow if the ground state were also symmetric. In particular, the force carriers gain masses. One way in which the symmetry can be broken is if the field related to the Higgs boson -- the so-called BEH field -- does not vanish in the ground state.  The weak force carriers are affected by their interactions with the BEH field, and this interaction slows them down. Moving at speeds lower than the speed of light - the consequence of interacting with the BEH field in the ground state -- is equivalent to giving them masses different from zero, and by that making the weak interactions short range.

A BEH field different from zero in the ground state of the Universe entails additional consequences and, in particular, further modifications of the predictions that would follow from unbroken symmetry. For example, the symmetry allows two Higgs particles to collide and produce two $Z$ bosons, but does not allow a single Higgs particle to decay into a pair of $Z$ bosons. But, once the ground state of the Universe breaks the symmetry, the latter is also allowed to occur. (Strictly speaking, the Higgs boson cannot decay into two $Z$ bosons because the sum of their masses is larger than the mass of the Higgs boson. The more precise statement is that the Higgs boson decays into a real $Z$ boson and a virtual one which produces a pair of fermions. We will continue to refer to this process, with some abuse of language, as ``Higgs decay into two $Z$ bosons".) Similarly, the symmetry allows two $Z$ bosons to collide and produce two Higgs bosons, but does not allow a single Higgs boson production from $Z$ boson fusion. But, once the ground state of the Universe breaks the symmetry, the latter process is also allowed to occur. Moreover, as stated above, the theory loses nothing of its predictive power. In fact, the strength of the interaction of the $Z$ boson with the {\it BEH field}, measured by the mass it gains from this interaction, is closely related to the strength of the  interaction of the $Z$ boson with the {\it Higgs particle}, measured by the rate at which the Higgs boson decays into two $Z$ bosons, and by the rate at which it is produced by $Z$ boson fusion.

The rate of the Higgs decay into two $Z$ bosons has been measured by the ATLAS and CMS experiments \cite{Sirunyan:2017exp,Aaboud:2017vzb}. Within present experimental accuracy, it has the value that corresponds to the strength of interaction that would give the $Z$ boson its mass, and via that would limit the effects of $Z$-mediated weak interaction to short range. Similarly, the rate at which the Higgs boson decays into a pair of $W$ bosons has been measured by the ATLAS and CMS experiments \cite{Khachatryan:2016vau}. Within present experimental accuracy, it has the value that corresponds to the strength of interaction that would give the $W$ boson its mass, and via that would limit the effects of $W$-mediated weak interaction to short range. Moreover, the rate at which a single Higgs boson is produced in vector boson fusion has also been measured by the ATLAS and CMS experiments and corresponds to the same strength of interaction.

Thus, these experiments have established a new law of Nature: The force carriers of the weak interaction gain their masses via their interaction with the everywhere-present BEH field, namely the field related to the newly discovered spin-0 particle. The strength of this interaction is precisely of the right size to limit the effects of the weak interaction to distances shorter than $10^{-18}$ meters.

\section{A third jewel: How the tau-lepton and the top- and bottom-quarks gain their masses}
The list of elementary matter (spin-1/2) particles known to us has twelve particles, divided into three groups or, in the physics jargon, generations. Each generation has four different particles: two particles that are sensitive to the strong interaction (``quarks") with electromagnetic charges (in units of the charge of the proton) $+2/3$ (up-quarks) and $-1/3$ (down quarks), and two particles that are insensitive to the strong interaction (``leptons") with electromagnetic charges $-1$ (charged leptons) and $0$ (neutrinos). The same symmetry that predicted that the weak-force carriers are massless, predicted also that all the matter particles known to us should be massless. Experiments have shown, however, that all matter particles are massive (with the one possible exception of the lightest neutrino). Is the symmetry principle challenged again? Not anymore: the fact that this symmetry is broken in the ground state of the Universe opens the door to the possibility that the matter particles would gain masses as well. But via what mechanism? This had been yet another open question that was answered by ATLAS and CMS.

As explained above, the LHC experiments have established that the way that the symmetry that predicts the existence of the weak interactions is broken in the ground state is by the BEH field. Can this force field, which slows down the $W$- and $Z$-bosons, also slow down the fermions? The answer is in the affirmative. But for this to happen, a new type of interaction has to exist: An interaction mediated by a spin-0 mediator, the Higgs boson itself. Such an interaction, called in the physics jargon ``Yukawa interaction",  has never been observed before among elementary particles. (As stated above, gravitational interactions are mediated by a spin-2 mediator, and the strong, electromagnetic and weak interactions by spin-1 mediators.) Discovering the Higgs decay into a pair of fermions would mean a discovery of this new type of interaction, the Yukawa interaction.

Moreover, as is the case for the weak force carriers, the strength of the interaction of a matter particle with the BEH field, measured by the mass it gains from this interaction, is closely related to the strength of the Yukawa interaction of that matter particle with the Higgs particle, measured by the rate at which the Higgs boson decays into two such fermions. (In the case of the top quark, the Higgs boson cannot decay into top-antitop pair because the sum of their masses is larger than the mass of the Higgs boson. To extract the strength of the Higgs-top interaction, experiments measure the rate at which the three particles -- a Higgs boson, a top quark and a top antiquark -- are produced.)

The three heaviest spin-1/2 particles -- the top quark, the bottom quark and the tau lepton -- are expected to have the strongest couplings to the Higgs boson, and consequently the largest rates of Yukawa interactions with it. The rate of the Higgs decays into the tau lepton-antilepton  pair has been measured by the ATLAS and CMS experiments \cite{Aad:2015zrx,Sirunyan:2017khh}. Within present experimental accuracy, it has the value that corresponds to the strength of interaction that would give the tau lepton its mass. The rate of the Higgs decays into the bottom quark-antiquark  pair has been measured by the ATLAS and CMS experiments \cite{Aaboud:2018zhk,Sirunyan:2018kst}. Within present experimental accuracy, it has the value that corresponds to the strength of interaction that would give the bottom quark its mass. The rate of the production of a Higgs boson together with a top quark-antiquark  pair has been measured by the ATLAS and CMS experiments \cite{Aaboud:2017jvq,Sirunyan:2018hoz}. Within present experimental accuracy, it has the value that corresponds to the strength of interaction that would give the top quark its mass.

Thus, these experiments have discovered a new fact about Nature: The third generation particles -- the tau lepton, the bottom quark and the top quark -- gain their masses via their interaction with the everywhere-present BEH field. This is also the discovery of a new, and rather special type of interaction: The Yukawa interaction, mediated by a spin-0 force carrier, the Higgs boson.

\section{Summary}
The LHC experiments, beyond constituting amazing intellectual and technological achievements, have made a series of profound discoveries about Nature and our Universe. While theorists had speculated about the existence of a spin-0 particle, whose non-zero force field is responsible for the short range of the weak interactions and for the masses of the spin-1/2 particles, long before the experiments discovered it, this must not diminish the significance of experimentally establishing these new laws of Nature.

Answering questions about Nature almost always leads to asking new questions. The discovery of the Higgs boson is the source of two new intriguing questions. First, the value of the Higgs mass implies that our Universe is likely to be in an unstable state. In the (far) future, a transition should happen to an entirely different Universe. Is it true that not only there is nothing special about Earth, or the solar system, or the Milky Way galaxy within the universe, but that in fact the fundamental structure of the entire Universe is only a temporary one?

Second, the lightness of the mass of the Higgs boson, compared to the Planck scale (above which quantum gravity effects become significant) or to the seesaw scale (below which new particles, beyond those of the Standard Model, must exist), poses a challenge to the very basic framework that we use in order to formulate the basic laws of Nature -- quantum field theory. In that framework, the lightness of the Higgs boson requires extreme fine-tuning (perhaps by as much as thirty two orders of magnitude) between seemingly unrelated constants of Nature. This would violate a principle called, in the physics jargon, ``naturalness". Is Nature in fact unnatural?

The potential of the LHC in discovering new facts about Nature and the Universe is far from saturated. There are at least two additional major open questions in our understanding of Nature that are guaranteed to be answered by the LHC experiments. First is the understanding of the mechanism that gives the second generation particles -- in particular, the muon and the charm quark -- their masses. (For recent progress, see \cite{Sirunyan:2020two,Aad:2020xfq}.) That may be the same mechanism as the one that has been shown to give the third generation fermion masses, or different from it. Second is the question of what happened at the electroweak phase transition, a cosmological event that took place when the Universe was about $10^{-11}$ second old. It may have been a smooth crossover, as predicted by the combination of the Standard Model of particle physics and the Big Bang model, or it may have been a first order phase transition, which would open the door for a new mechanism to explain the matter-antimatter imbalance in the Universe.

The LHC experiments have already three jewels -- major discoveries -- in their crown, and there are additional jewels -- major discoveries -- guaranteed to be added to it.

\vskip1.0cm
\noindent{\bf Acknowledgments:}
I thank Kfir Blum and Gilad Perez for helpful discussions and comments on the manuscript.
YN is the Amos de-Shalit chair of theoretical physics, and is supported by grants from the Israel Science Foundation (grant number 1124/20), the United States-Israel Binational Science Foundation (BSF), Jerusalem, Israel (grant number 2018257), and by the Minerva Foundation with funding from the Federal Ministry for Education and Research).


\end{document}